# Wafer-scale Programmed Assembly

# of One-atom-thick Crystals

*Seong-Jun Yang[1,†], Ju-Hyun Jung[1,†], Eunsook Lee[2], Edmund Han[3], Min-Yeong Choi[1], Daesung Jung[4], Shinyoung Choi[1], Jun-Ho Park[1], Dongseok Oh[2], Siwoo Noh[2], Ki-Jeong Kim[2], Pinshane Y. Huang[3], Chan-Cuk Hwang[2], Cheol-Joo Kim[1,*]*

[1]Department of Chemical Engineering, Pohang University of Science and Technology, Pohang, Gyeongbuk 37673, Republic of Korea.

[2]Beamline Research Division, Pohang Accelerator Laboratory, Pohang, Gyeongbuk 37673, Republic of Korea.

[3]Department of Materials Science and Engineering, University of Illinois at Urbana-Champaign, Urbana, Illinois 61801, USA.

[4]Convergence Research Center for Energy and Environmental Sciences, Sungkyunkwan University, Suwon, Gyeonggi 16419, Republic of Korea

* Correspondence to: kimcj@postech.ac.kr

† These authors contributed equally: Seong-Jun Yang, Ju-Hyun Jung

ABSTRACT

Crystalline films offer various physical properties based on the modulation of their thicknesses and atomic structures. The layer-by-layer assembly of atomically thin crystals provides powerful means to arbitrarily design films at the atomic-level, which are unattainable with existing growth technologies. However, atomically-clean assembly of the materials with high scalability and reproducibility remains challenging. We report programmed crystal assembly (PCA) of graphene and monolayer hexagonal boron nitride (ML hBN), assisted by van der Waals interactions, to form wafer-scale films of pristine interfaces with near-unity yield. The atomic configurations of the films are tailored with layer-resolved compositions and in-plane crystalline orientations. We demonstrate batch-fabricated tunnel device arrays with modulation of the resistance over orders of magnitude by thickness-control of the hBN barrier with single-atom precision, and large-scale, twisted multilayer graphene with programmable electronic band structures and crystal symmetries. Our results constitute an important development in the artificial design of large-scale films.

Since the reports of various two-dimensional (2D) materials after the isolation of graphene, assembly of the materials into artificial structures has been intensively studied for discovery of novel properties and fabrication of advanced nano-devices[1]. The assembly of 2D materials provides a variety of van der Waals structures by structural degrees of freedom that are difficult to achieve with conventional deposition techniques of bulk materials or thin films that exploit self-assembly processes, and has two unique advantages. First, the separation of the growth and assembly steps enables various combinations of materials that have incompatible growth conditions[2,3]. Second, the absence of chemical bonds between adjacent layers allows for control of the crystalline orientation of individual layers to modulate atomic configurations of the whole structure[4,5]. Therefore, even with small varieties of 2D materials, numerous multilayer structures have been fabricated to host programmed interactions with various fundamental particles, including electrons[6,7,8], ions[9], phonons[10], and polaritons[11].

Graphene and hBN form the key assembly units with unique electrical, optical and mechanical properties. Mechanical assembly of 2D materials, assisted by van der Waals interactions[2,12,13] produces pristine, controlled interfaces through which the properties of final stacks are programmed. However, current mechanical methods to assemble graphene and hBN have low throughput, due to the lack of reliable production of uniform assembly units. They are obtained either by exfoliating flakes from bulk crystals, or by using chemical vapor deposition (CVD) to grow films. Exfoliated flakes have spatially non-uniform thickness and small sizes, typically on the scale of tens of micrometers, limited by the size of bulk crystals[12,13]. Epitaxial growth by CVD produces wafer-scale films with aligned crystallinity[14,15], but the conventional process to assemble the films involves wet-process, which induces interfacial contaminants[16]. In the last few years, there has been rapid progress in the assembly of large-scale transition metal dichalcogenides[2], however atomically clean assembly of graphene and hBN crystals is still elusive, and remained as a key technique to be developed (see supporting text and fig. S1). The reliable assembly of graphene and hBN with atomic-level precision would enable fabrication of high-quality, van der Waals structures with greatly extended tunability. Realizing this approach is paramount for the discovery of new properties and the development of devices based on 2D materials.

Here, we report programmed crystal assemblies of graphene and ML hBN films to fabricate high-quality, spatially-uniform multilayer films. Two examples are shown to represent the main capabilities of our assembly technique. The first example is a graphene/hBN vertical superlattice, fabricated by alternately stacking graphene and ML hBN to a total number of layers $N_L$ = 5 (Figure 1a). The cross-sectional scanning transmission electron microscopy (STEM) image (Figure 1b) showed five parallel layers with pristine interfaces, with interlayer spacing $d_z$ = 3.37 Å (fig. S2), which is close to the expected value for graphene and hBN[17]. Electron energy loss spectroscopy (EELS) revealed vertical chemical compositions with alternating C and N peaks along the out-of-plane direction (Figure 1c). Significantly, we observe no additional signals of contaminant elements, such as amorphous hydrocarbon.

The second example is chiral twisted graphite (CTG) with $N_L$ progressively-twisting layers around the stacking direction with a constant interlayer rotation $\theta_i$ (Figure 1d). A top-down TEM diffraction pattern of a chiral stack with $N_L$ = 6 and $\theta_i$ = 10° (Figure 1e) showed diffraction spots with rotational periodicity of ~10°, which indicates angle alignment within ±1°. The atomic structure of a chiral film was directly imaged by cross-sectional STEM on a sample with $N_L$ = 10 and $\theta_i$ = 20° (Figure 1f). The graphene layers were visible as 10 bright lines with a constant $d_z$ = 3.41 Å (fig. S2). Three layers (arrows), separated by $3d_z$ exhibit higher intensity than the surrounding layers and lattice fringes along the in-plane direction. Analysis of a fast Fourier transform (FFT) of the image (Figure 1g) identified that the lattice fringes show a periodicity of 2.15 Å, which corresponds to the graphene lattice constant projected along the armchair $\langle\bar{1}110\rangle$ direction $d_x$. The vertical three-layer periodicity is expected for the $\theta_i$ = 20° stacking configuration: graphene has six-fold symmetry and should reach an equivalent orientation every three layers (20°/layer × 3 layers = 60°). The increased intensity may be attributed to electron channelling, which occurs when the electron beam is oriented near a high-symmetry zone axis. Collectively, these results (Figure 1) demonstrate that both chemical compositions and atomic arrangements of van der Waals structures can be controlled at the atomic-level by PCA of graphene and ML hBN crystals.

Our PCA for atomic-scale topography involves four general steps (Figure 2a; details in supporting methods): (I) growth of high-quality graphene and ML hBN films with uniform thickness and crystalline orientation on Ge(110) single-crystalline substrates by CVD[14,18], (II) mechanical exfoliation of a film (graphene or ML hBN) using thermal release tape after Au superlayer deposition, (III) placement of exfoliated film on top of another as-grown film at angle $\theta_i$ (steps II and III are performed $N_L$ times), and (IV) transfer of stacked film onto a target substrate.

Epitaxial graphene and ML hBN grown on Ge(110) serve as ideal structural units to realize a scalable and atomically-clean assembly process. After optimal growth (fig. S3), graphene and ML hBN showed ML thickness with uniformity > 99 %, low atomic defects (fig. S4) and single crystallographic orientation across the area of a 2" wafer, as confirmed by low-energy electron diffraction (LEED) (Figure 2b, fig. S5). The epitaxial films have straight-edge cuts (fig. S6) with certain crystalline orientations determined by the Ge(110) epi-substrates, which enable control of $\theta_i$ without additional characterization of the crystalline structures. Epitaxially-grown single-crystalline films, as previously grown on metals, are usually strongly bound to them, so they must be removed by chemical etching to isolate the films. Importantly, the interaction energy between the Ge(110) substrate and the as-grown films is lower than between layered materials[19], so efficient mechanical exfoliation and stacking can be performed by exploiting van der Waals interactions without exposing the interfaces to other substances, such as polymer or etchant[16].

Multilayer films fabricated by the assembly method were uniform over large-scale. To demonstrate the scalability, wafer-scale graphene films (Figure 2c) were consecutively stacked, and the stacked film was transferred onto a $SiO_2$/Si substrate. To achieve efficient exfoliation, each film was larger than the one below it. The optical spectra and image (Figure 2d, fig. S7) of the films showed spatially uniform contrasts, which increased linearly as the $N_L$ increased. Near-unity yield of stacking over a macroscopic area was also confirmed by optical absorption measurements in hBN multi-stacks (fig. S7). The multilayer films offer excellent structural integrity with minimal defects, such as cracks, and therefore yielded suspended films on a holey grid (fig. S8).

The quality of interfaces was investigated in multilayer stacks by conducting out-of-plane X-ray diffraction (XRD) on a millimeter scale. The representative diffraction peak, measured in a CTG with $\theta_i$ = 20° and $N_L$ = 10 over a 0.6 cm² area (Figure 2e, orange line) showed significantly stronger intensity and lower full width at half maximum than in the reference multilayer film (gray line), prepared by wet-transfer. The $d_z$ of 3.42 Å, deduced from the peak position is consistent with the value, measured by direct imaging in STEM (Figure 1f and 2h). Based on the Scherrer equation (supporting text), the thickness of coherently reflecting lattices is estimated as ~3.6 nm, nearly the ($N_L$ - 1) × $d_z$, strong evidence of atomically clean interfaces with uniform $d_z$ across the millimeter scale sample size. The $d_z$ were also measured for CTG with different $\theta_i$ (Figure 2f) to check the reliability of our PCA process for arbitrary twist angle. At $\theta_i$ = 0°, $d_z$ was measured as 3.34 Å, which is close to the value for the natural graphite. $d_z$ increased by 2 % to 3.42 Å as $\theta_i$ increased to 3°, then shows almost constant values for $\theta_i$ > 3°. This dependence of $d_z$ on $\theta_i$ is consistent with a theoretical model[20] (gray line), which considers the structural relaxation by the interlayer interactions. As $\theta_i$ increases, Bernal-stacked regions transform to incommensurate structures with high local $d_z$, so the average $d_z$ increases until the transformation is complete near $\theta_i$ = 3.3°[5]. We further examined the cleanliness and surface morphology of multilayers by atomic force microscopy (Figure 2g, fig. S9 and fig. S10), X-ray photoelectron spectroscopy (fig. S10), and optical hyperspectral imaging (fig. S11). Collectively, our results demonstrate that the PCA process forms pristine interfaces.

Our PCA is a powerful tool to generate functional devices, in which atomic-scale control of the structures with pristine interfaces is crucial to obtain desirable properties. We demonstrated the versatility and strength of PCA by producing batch-fabricated tunnel device arrays. They are composed of 2D materials and metallic grids, in which vertical charge transport is strongly affected by interfacial structures such as thickness, atomic defects, and doping of constituent layers[21]. Here, centimeter-scale graphene film was assembled with hBN with controlled $N_L$, then transferred onto gold electrodes to form an array of metal-insulator-metal junctions (Figure 3a). The current densities $J$ normalized by the junction area were measured in ambient conditions. $J$ decreased significantly by ~1 % per layer as $N_L$ increased (Figure 3b). Zero-bias resistances $R_0$ multiplied by junction area $A$ (Figure 3b, inset) were measured from 22 devices and plotted in the Figure 3b inset. We fitted the $N_L$-dependent $R_0A$ with a basic tunneling equation by taking the barrier height as a free parameter on the assumption that hBN layers have ideal thicknesses with pristine interfaces (supporting text). The barrier height was deduced to 3.0 eV for hole-tunneling current; this height is consistent with the energy difference between Fermi level and valance band maximum of hBN on graphite, measured by angle-resolved photoemission spectroscopy[22] (ARPES).

The tunneling current is also determined by the density of states of constituent elements. The plot of differential conductance d$I$/d$V$ vs bias voltage $V$ (Figure 3d) shows a dip near the zero-bias region by suppression of phonon-mediated tunneling[23], plus two local minima at biases of $V_A$ and $V_D$. Local minima develop when the Fermi level of either electrode is aligned to the Dirac point of graphene that has the minimum density of states (Figure 3c; fig. S12 for confirmation of band alignments). The simulated d$I$/d$V$ curve (Figure 3d inset) obtained using a phonon-mediated tunneling model for hole-transports (fig. S13) agrees well with the experimental data; this result confirms the formation of pristine interface with minimized defect-mediated tunneling.

The PCA method also allows scalable and clean integration of various 2D materials, and thereby offers a general approach to design novel van der Waals structures. As an example, as-grown $MoS_2$ films were integrated with graphene at near-unity yield to form a uniform graphene/$MoS_2$ superlattice (fig. S14). The result indicates that all key electronic elements, including metal, semiconductor, and insulator can be assembled with high-quality interfaces on a large-scale. Furthermore, with twist at the pristine interfaces, $\theta_i$-dependent interlayer interactions occur, and determine the topology and symmetry of electronic bands[24]. Therefore, $\theta_i$ control at multiple interfaces realizes increases the variety of properties for the given material platform.

We demonstrated a structurally-tunable van der Waals platform by fabricating two types of twisted graphite with programmed band topologies (Figure 4b, c) that have a single value of $|\theta_i|$, but different combinations of rotational polarity. Graphene layers stacked with a constant $\theta_i$ toward the same rotational polarity produce CTG with periodic electronic bands in the momentum space, hosting equally-spaced hybrid states with mini-gaps by interlayer interactions (Figure 4b, right; red-colored lines). In contrast, stacking of $\theta_i$ of alternating sign with an odd $N_L$ forms achiral twisted graphite (ATG) (Figure 4c, left) with a mirror plane in the middle layer. In ATG, the hybrid states of constituent interfaces are superimposed at the same momentum space to interact with each other, so they generate collective states of different topology[25–28].

ARPES over the large-area (100 μm × 300 μm, Figure 4a) revealed the band structures of CTG and ATG ($N_L$ = 3, $|\theta|$ = 20°). Energy dispersion cross-sections were obtained along the multiple Dirac points and hybridization points (HPs, defined in fig. S15) in the momentum space (Figure 4d, e). In CTG (Figure 4f), three equally spaced, conical bands with gradually attenuating intensities form mini-gaps at HPs at constant energy (red arrows) and parabolic bands (red dotted lines) (fig. S15-17 for additional data); this result represents hybridization between bands from adjacent layers. In ATG (Figure 4g), the interlayer hybridizations also happen near the crossing point between $L_2$-band and superimposed $L_1/L_3$-bands to yield the emergent parabolic band (red dotted line). The confirmation of band structure engineering suggests that PCA can realize the numerous electronic states and correlated properties that have been proposed for ATG[25–28] and CTG[4,29] with effective interlayer interactions across pristine interfaces. Fabrication of twisted multilayer structures with programmed band structures has been technically challenging, because even small amounts of contaminants during the stacking can prevent the interlayer interactions. Therefore, most studies on twisted multilayer crystals have been conducted in samples with small $N_L$. Our assembly provides atomically clean interfaces with near-unity yield, therefore one can reliably program band structures in multilayer films even with large $N_L$, where various properties are theoretically predicted[28,29] (fig. S18 for chiro-optical property as an example).

Our PCA method can control not only the size, but also the atomic shape of the materials to program their intrinsic properties, heralding an exciting era of 'nano-topotechnology'. Numerous artificial structures can be fabricated by a designer approach with single-atom precision on wafer-scale to study interactions with fundamental elements, facilitating the discovery of various physical properties. For example, van der Waals crystals with controlled interfaces can result in strong infrared photoresponse[30], ultrathin thermal isolation[31], enhanced superconductivity[32] and fast ion diffusions at van der Waals interfaces[33], which are distinct from natural crystals. In particular, since associated characteristic length scales are very small, such as phonon wavelengths[34] and the radii of interlayer excitons[35], the properties strongly depend on the interface structures at the atomic scale. Due to the difficulty of forming pristine interfaces, these properties have been demonstrated mostly in samples with a limited number of layers and lateral sizes. Our reliable and scalable method provides a powerful tool for investigating more complex structures and accelerating the development of advanced devices based on van der Waals crystals.

ASSOCIATED CONTENT

**Supporting Information**. The Supporting Information includes following contents.

Experimental details for growth of 2D materials, PCA and wet process of 2D materials, device fabrication and other characterization tools. Discussions of 2D assembly developments, XRD for twisted graphite, ARPES data from tri-layer CTG and ATG, circular dichroism for CTG, tunneling model for Gr/hBN/Au tunnel junctions, estimation of tunneling barrier height. Supporting figures for summary of 2D assembly developments, cross-sectional STEM, graphene growth, twist angle-controlled stacking, near-unity stacking yield, mechanical robustness of multilayer membrane, surface flatness and cleanliness of multilayer graphene, interlayer optical absorption of twisted bilayer graphene film, Gr/hBN/Au junction devices, graphene/$MoS_2$ superlattices, band structure of CTG and ATG and chiro-optical properties in CTG.

AUTHOR INFORMATION

**Corresponding Author**

*E-mail: kimcj@postech.ac.kr

**Author Contributions**

S.-J.Y. and C.-J.K. designed the experiments. S.-J.Y., J.-H.J. and M.-Y.C. synthesized the 2D films. S.-J.Y. and J.-H.J. conducted the assembly process. E.H. and P.Y.H. performed atomic-resolution TEM imaging. E.L., D.J., D.O. and C.-C.H. conducted XPS and ARPES measurements. J.-H.J., S.N. and K.-J.K. conducted LEED measurements. S.C. and J.-H.P. performed hyperspectral imaging. S.-J.Y. and J.-H.J. performed other sample characterizations. S.-J.Y. and C.-J.K. wrote the manuscript. All authors discussed the results and commented on the manuscript.

**Notes**

The authors declare no competing financial interest.

ACKNOWLEDGMENT

This work was supported by the Basic Science Research Program (2020R1C1C1014590), the Basic Research Laboratory Program (NRF-2020R1A4A1019455) and the Creative Materials Discovery Program (NRF-2018M3D1A1058793, NRF-2020M3D1A1110548) of the National Research Foundation of Korea (NRF) funded by the Korea government (Ministry of Science and ICT). The electron microscopy work was supported by NSF-MRSEC award no. DMR-1720633, using the Materials Research Laboratory Central Facilities at the University of Illinois, where electron microscopy support was provided by J. Mabon, C. Chen and H. Zhou. C.-C. Hwang was supported by the NRF funded by the Ministry of Science and ICT (2020R1A2B5B02001876, 2018R1A56075964).

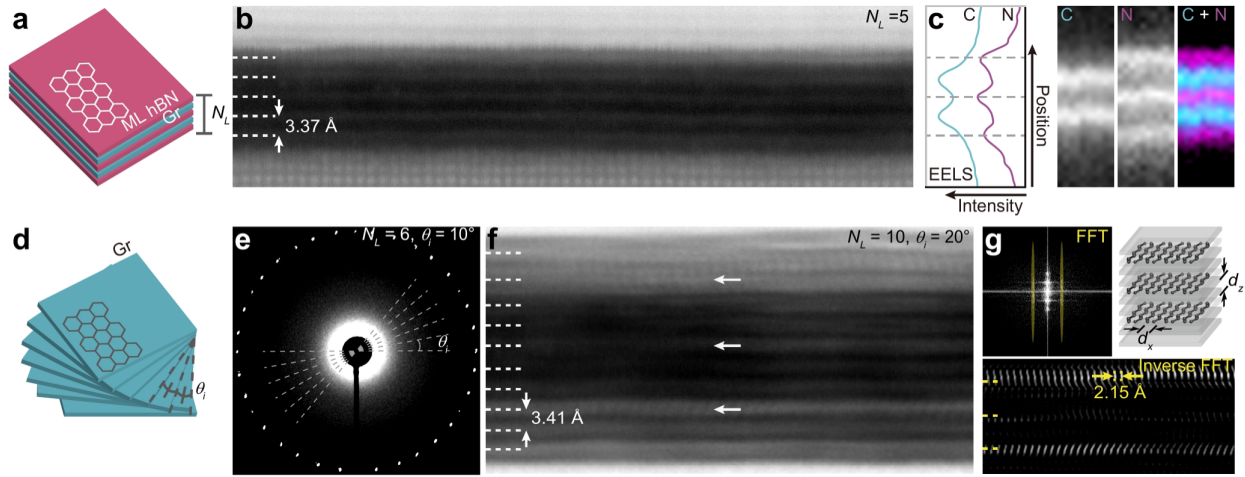

**Figure. 1. Artificial van der Waals films with high-quality interfaces.** (a) Schematic of graphene/hBN vertical superlattice. (b) Annular dark-field STEM image of cross-section of the graphene/hBN vertical superlattice ($N_L$ = 5). (c) EELS intensity profiles and composition maps along out-of-plane direction. The apparent curvatures of layers in the EELS maps are a result of specimen drift during acquisition rather than real sample roughness. (d) Schematic of CTG. (e) TEM diffraction pattern from a top view of a CTG ($N_L$ = 6, $\theta_i$ = 10°) that is measured in > 400 nm in diameter. (f) Annular dark-field STEM image of cross-section of CTG ($N_L$ = 10, $\theta_i$ = 20°). (g) FFT image of the graphene region in (f) (top, left). Inverse FFT image of the STEM image, corresponding to the encircled spots in the FFT pattern (bottom). Schematic of the atomic structures of the CTG (top, right).

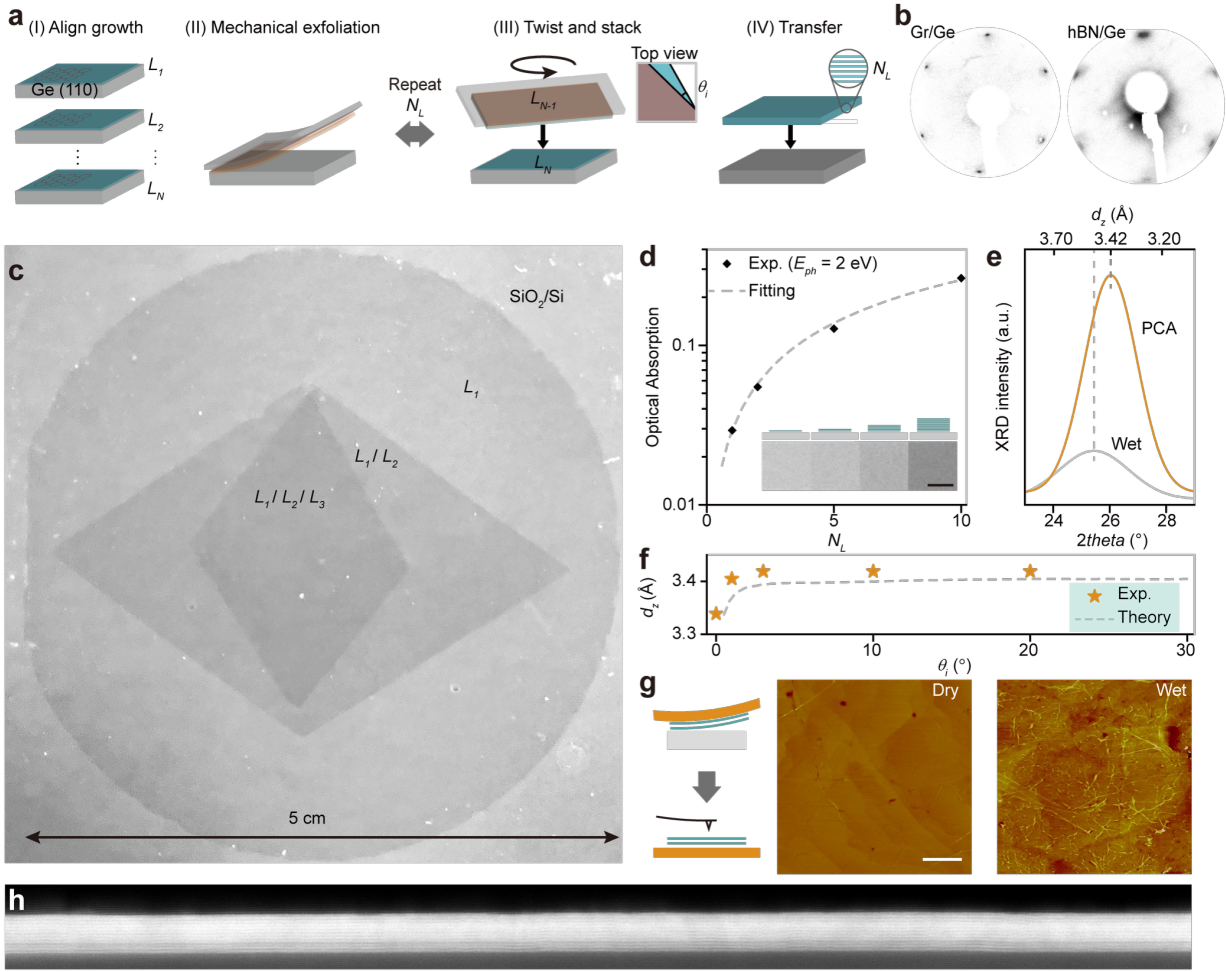

**Figure. 2. Wafer-scale PCA process for atomic-scale structural engineering.** (a) Schematics of assembly of one-atom-thick crystals, which is conducted in all-dry conditions by exploiting van der Waals interactions. (b) LEED pattern of as-grown graphene (left) and hBN (right) on Ge(110) substrates. (c) A grayscale optical image of wafer-scale, assembled graphene films on a SiO₂/Si substrate. (d) Optical absorption values at 2 eV for films with different $N_L$, and the expected trend (dashed line) $1 - (T_{SLG})^{N_L}$, where $T_{SLG}$ is the reported transmittance of single-layer graphene. (Inset: optical images of samples with different $N_L$. Scale bar: 1 mm.) (e) XRD data for films ($N_L = 10$, $\theta_t = 20°$), prepared by PCA (orange line) and wet-stacking processes (gray line). (f) Measured $d_z$ by XRD in CTG with different $\theta_t$ (orange stars) and theoretically

predicted values (gray dashed line)[18]. (g) Measurement schematics (left) and AFM images of the bottom surfaces of bilayer films, which are prepared by all-dry stacking (center) and wet stacking (right) processes, respectively. The dry stacking sample shows a much flatter surface with fewer wrinkles. (Scale bar: 1 μm, height scale: 50 nm.) (h) Cross-sectional bright-field STEM image of the film in (e) by PCA across 100 nm.

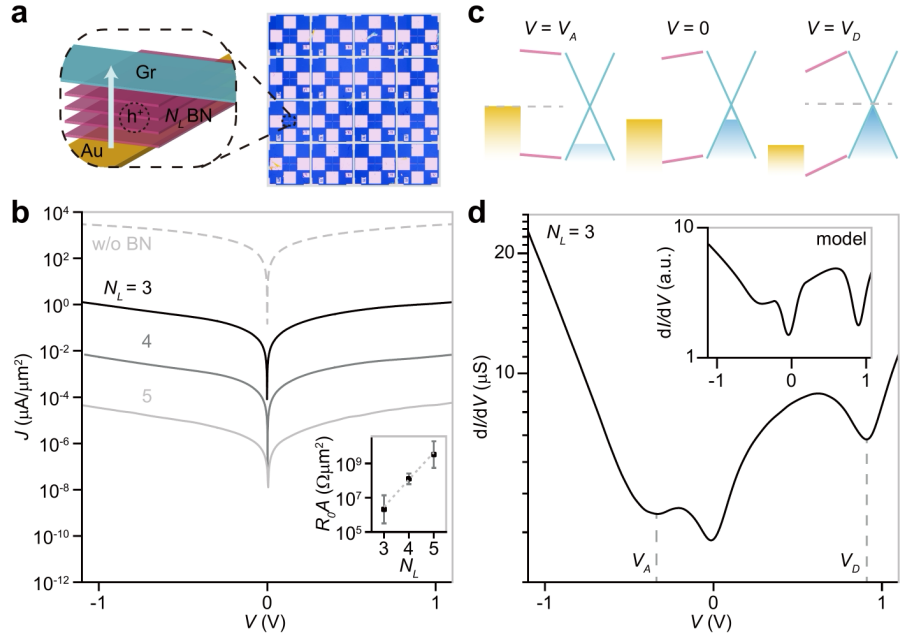

**Figure. 3. Batch-fabricated tunnel devices.** (a) Schematic (left) and optical image (right) of batch-fabricated Gr/hBN/Au tunnel junctions by PCA. (b) $J$-$V$ curves for hBN tunneling barriers with different $N_L$. (Inset: $N_L$-dependent $R_0A$, averaged from multiple devices with a fitting (dashed line) by a tunneling model.) (c) Schematics of band diagrams at different $V$, for local minima of d$I$/d$V$ (d) d$I$/d$V$-$V$ curve for $N_L$ = 3. (Inset: Simulated curve with a tunneling model.)

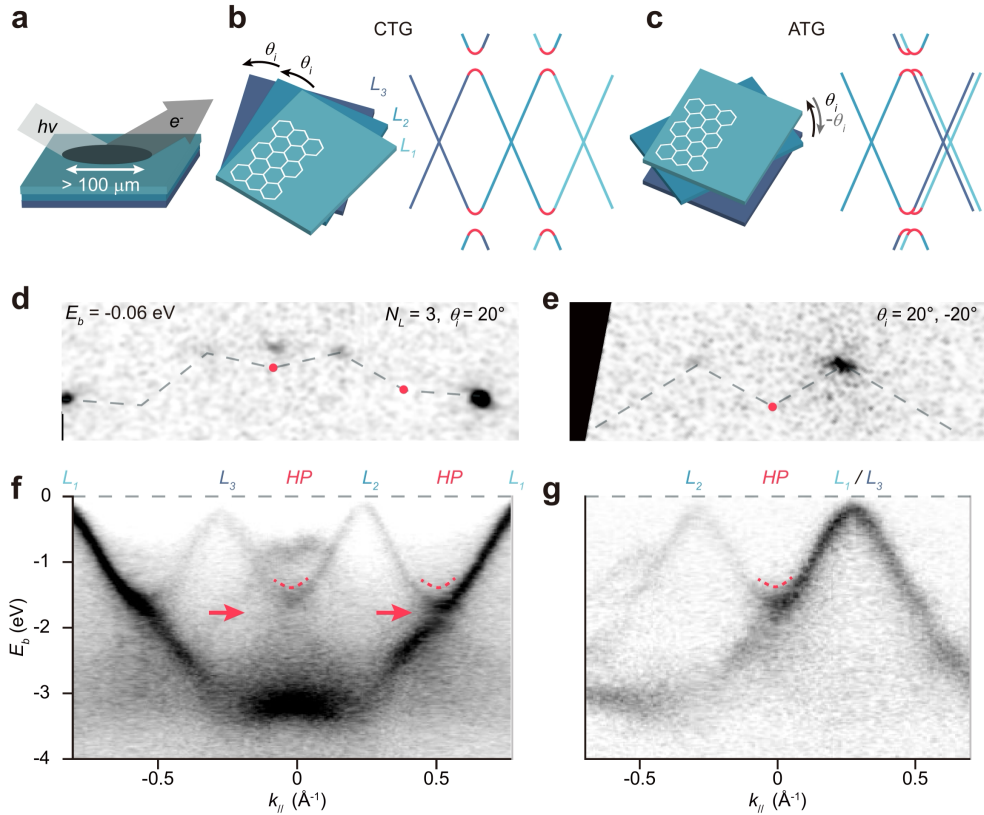

**Figure. 4. Twisted graphite with programmable electronic band structures.** (a) Schematic for large-scale ARPES measurement. (b, c) Schematics for the stacking configuration and electronic band structures of CTG with a constant rotational polarity (b) and ATG with alternating polarities (c). (d, e) Photoemission contours at $E_b$ of –0.06 eV for CTG ($N_L$ = 3, $\theta_i$ = 20°; d) and ATG ($N_L$ = 3, $\theta_i$ = ±20°; e). (f, g) Band structures for the CTG (f) and ATG (g) along dashed lines in the contour maps (d) for CTG and (e) for ATG. Dirac cones from the nearest layers form hybridized states at HPs. Parabolic bands (red dotted lines) with mini-gaps (red arrows) appear at HPs.

# Supporting Information

## Contents

### Supporting Methods

Growth of 2D materials

PCA process

Wet assembly

Device fabrication

Device measurement

Other characterizations

### Supporting Text

Summary of 2D assembly developments

XRD from twisted graphite

ARPES data from tri-layer CTG and ATG

Circular dichroism from multilayer CTG

Tunneling model for vertical transports across Gr/hBN/Au junctions

Estimation of tunneling barrier height

## Supporting Methods
### Growth of 2D films

Graphene: ML films were grown on single-crystalline Ge(110) substrates (AXT, ingot #: GH221 and Umicore, product #: 96462) by CVD[14]. First, the Ge substrates were sonicated sequentially in acetone and isopropyl alcohol for 5 min each, to remove organic residues. The substrates were then dipped into 20 % HF solution for 5 min to etch away any native oxides. After loading substrates, a quartz tube furnace was evacuated to 100 mTorr, then filled with Ar (200 sccm) and $H_2$ (20 sccm) to ambient pressure. Graphene films were grown for 13.5 h at 923 °C with 1 % $CH_4$ in $H_2$ (6 sccm), $H_2$ (3 sccm) ($CH_4$/$H_2$: 0.067), and Ar (200 sccm).

hBN: ML films were grown by CVD[18] on Ge(110) substrates that had been pre-treated as for fabrication of graphene films. After loading the substrates, the furnace was evacuated, then filled with $H_2$ (200 sccm) to ambient pressure. The hBN films were grown for 17 h at 926 °C with $H_2$ (200 sccm) and borazine (stored at −20 °C), carried by Ar (0.0003 sccm).

$MoS_2$: ML films were grown on fused silica substrates by using metal-organic CVD[36]. The furnace was evacuated to 50 mTorr, then Ar (150 sccm) and $H_2$ (10 sccm) were subsequently introduced while ramping up the temperature to 580 °C. Diethyl sulfide (1 sccm) was initially introduced for 30 min, then both molybdenum hexacarbonyl (0.2 sccm) and diethyl sulfide (1 sccm) were flown at 9.5 Torr for 18 h.

### Programmed crystal assembly

(1) Preparation of 1st stack layer: A 40-nm-thick Au film was deposited on as-grown graphene/Ge(110) or hBN/Ge(110) substrates by thermal evaporation to serve as a mechanical support for 2D film. Polymethyl methacrylate (PMMA, 996 K, 8 % in anisole) was spin-coated over the Au film at 2000 rpm for 1 min, then baked at 180 °C for 10 min to introduce a soft layer for conformal contact during the stacking process. A thermal release tape (TRT) (Nitto, product #: 3195M) was attached to the surface as a handling frame that was used to mechanically peel the underlying PMMA/Au/2D film from the Ge(110) substrate.

(2) Stacking and peeling: $\theta_i$ was controlled using a home-built transfer system equipped with a rotatable stage and an optical microscope. An as-grown graphene/Ge(110) substrate, with the graphene film facing up, was placed on the stage, which was heated to 80 °C. The 1st stacking layer was mounted on the micro-manipulator with the graphene surface facing down. We controlled $\theta_i$ by optically confirming the rotational angle between straight edges (fig. S6) of each layer. While the pick-up layer was gently pressed onto the next stack layer, they were annealed at 120 °C for 10 min to achieve conformal contact. The TRT was removed by annealing at 180 °C for 10 min. A new TRT was attached and used to mechanically peel the underlying layers from the bottom Ge substrate. The stacking and peeling processes were repeated to reach the target $N_L$. For assembly of composite van der Waals films, the graphene layers were alternately stacked with $MoS_2$ films or hBN films; other processes were the same

   (3) Transfer: The final multilayer films can be transferred onto arbitrary substrates. The stacked film was placed on a target surface then baked at 145 °C to release the TRT. The film was then annealed at 180 °C to promote interfacial adhesion between the graphene and substrates. The PMMA layer was removed by either dipping the substrates in chloroform for 4 h or baking at 315 °C for 3 h and further treated with $O_2$ plasma for 2 min to remove any residue. The Au film was etched with $KI/I_2$ solution, and the surface was rinsed with ethanol and deionized (DI) water.

   We note that it is important to use Au as the supporting material, because of several key characteristics of Au. Evaporated Au weakly interacts with graphene and hBN, therefore it can be gently and cleanly removed after transfers. Also, soft Au layer enables conformal contacts with the underlying substrates during the assembly and final transfer, which is important for the high-yield assembly without damaging the films.

**Wet assembly**
   PMMA was spin-coated onto an as-grown graphene/Ge(110) substrate, then baked at 180 °C for 10 min. It was immersed in 1 M NaOH aqueous solution, then PMMA/graphene film was electrochemically delaminated from the Ge substrate by generating $H_2$ bubbles at the interface between graphene and the Ge substrate with a current of 0.1 A across the Ge anode and PMMA/graphene/Ge cathode[37]. The film was rinsed in DI water five times, then transferred onto as-grown graphene/Ge(110). Before the PMMA was removed, the process was repeated to achieve the desired $N_L$.

**TEM**
   (1) Sample preparation: A protective layer of amorphous carbon (5~30 nm thick) was evaporated on top of the Au/CTG or Au/graphene/hBN superlattice film. Samples for cross-sectional TEM samples were prepared using standard focused ion beam lift-out procedures in a Thermo Fisher

Scientific Helios 600i Dual Beam FIB-SEM system. Final milling was performed at 2 kV using a cryo-can to reduce sample damage and minimize re-deposition.

(2) STEM and EELS imaging: Imaging was performed on the Thermo Fisher Scientific Themis Z aberration-corrected electron microscope at 80 kV, below the knock-on damage threshold of graphene and hBN. A convergence angle of 25.2 mrad was used. EELS experiments were conducted on the Gatan spectrometer, GIF Quantum 965. The full width at half maximum of the zero loss peak is 0.9 eV. EELS spectral images were acquired using a 5 mm aperture, dispersion of 0.1 eV/channel, pixel size of 0.98 Å, and a dwell time of 1 s/pixel for the core-loss measurements.

## LEED

All LEED patterns were taken in an ultra-high vacuum chamber at $10^{-9}$ Torr over ~1 mm$^2$ area using the ErLEED 150 optics system (Specs) with incident photons of 70~75 eV.

## XRD

Out-of-plane XRD was measured using a Bruker D2 PHASER with a beam size of ~1 cm$^2$. A *theta*-2*theta* scan of the sample was performed with Cu Kα radiation ($\lambda$ = 1.5406 Å, power: 300 W).

## ARPES, XPS

ARPES experiments were performed at the 10D beamline in Pohang Accelerator Laboratory equipped with a Scienta DA30 electron analyzer. Various photon energies were tuned to optimize photon energy for ARPES measurements. All ARPES data were taken with a photon energy of 55 eV. The samples were heated by electron bombardment for 1 min at 500 °C under ultra-high vacuum to eliminate surface contaminants during sample transportation in the air. All photoemission data were recorded at $1.0 \times 10^{-10}$ Torr and room temperature. XPS measurements were conducted using the same set-up used for ARPES. Photon energy was 360 eV.

## Optical absorption spectroscopy

The optical absorption spectra of films, transferred onto fused silica substrates, were measured in ambient conditions using transmission mode. The absorption was determined as $(1 - I/I_0) \cdot (n_s + 1)/2$, where $I$ and $I_0$ are respectively the transmitted light intensities of the substrate region with and without films, and $n_s$ is the refractive index of the substrate[38]. The absorption is equivalent to $1 - (T_{ML})^{N_L}$, where $T_{ML}$ is the transmittance for ML. We deduced the absorption $A_{ML}$ for ML by taking $1 - T_{ML}$ from measurements using multilayer samples with different $N_L$ (fig. S7C, D). For the local optical spectroscopy and spectral imaging in fig. S11, we used a home-built hyperspectral imaging set-up with a sub-micrometer spatial resolution[39].

**Circular dichroism measurement**

The optical circular dichroism was measured using a Jasco circular dichroism spectrometer (J-815) in transmission mode with lights, perpendicular to the film planes, over an area of 2 mm in diameter. Lock-in measurements were conducted by alternating at 50 kHz between left-handed and right-handed circularly polarized lights.

**Fabrications of tunnel devices**

Vertical tunneling devices were fabricated using conventional photolithography. First, to fabricate the bottom electrodes, photoresist lift-off was patterned on $SiO_2$/Si substrates. Then an Au (20 nm)/Cr (5 nm) bilayer electrode was deposited twice with a thermal evaporator in a high vacuum condition ($\sim 10^{-6}$ Torr). After lift-off, the substrates were treated with piranha solution for 7 min to remove any organic residues on the surface of patterned electrodes. The assembled PMMA/Au/ML hBN/Gr/$N_L$ layers of hBN film was transferred onto the electrodes, followed by PMMA dissolving with chloroform. The top Au/ML hBN/Gr/$N_L$ layers of hBN film was patterned by consecutive photolithography into stripes to form cross-bar arrays with the bottom Au/Cr electrodes, Au etching with $KI/I_2$ solution and graphene etching by $O_2$ plasma. Finally, the Au stripes masking graphene were removed by additional $KI/I_2$, resulting in graphene/$N_L$ layers of hBN/Au junctions.

**Tunneling current measurement**

All the tunneling currents were measured across the graphene/$N_L$ layers of hBN/Au junctions at ambient conditions. We applied superimposed DC and small AC ($V_{rms}$ = 2 mV) bias to the bottom Au electrode with respect to the top graphene electrode. While sweeping the DC bias voltage, the current density and differential conductance were simultaneously measured with a low-noise current preamplifier (DL instruments, 1211) and lock-in amplifier (Stanford research instruments, SR830). The current density across the junction was defined by dividing the current by the junction area of $\sim 10 \ \mu m^2$ between two electrodes.

The electrical properties of graphene electrodes were separately investigated by measuring the lateral charge transports with a source-drain bias, applied along the graphene (fig. S12A). The bottom Au electrode underneath hBN can be used as a back gate to obtain transconductance data of the graphene channel at different gate-bias conditions. For the measurements, the source-drain voltage was set to be high enough to induce sufficiently large lateral currents to minimize the contribution from the vertical tunneling current through hBN tunneling barriers. As a result, the transconductance curve (fig. S12B) shows Dirac point resistance with the maximum value near the gate voltage of $V_D$, as defined in Figure 3d.

## Supporting Text

### Summary of the developments for 2D crystal assembly

Our result represents an important step towards developing the material design tool based on assembly of two-dimensional materials by resolving the issue of the limited cleanness at the interfaces of graphene and hBN. The schematic visualization of the technology roadmap is shown in fig. S1, where we classified the developments into different categories of (i) exfoliated crystals, (ii) CVD-grown transition-metal dichalcogenides (TMD) and (iii) CVD-grown graphene and hBN, depending on the types of assembly units. The full development with a high production yield can be realized with the 3 key techniques for (1) production of crystalline units, (2) atomically clean assembly, and (3) automated assembly. After the first isolation of graphene[40], the assembly of two-dimensional materials was started with exfoliated crystals. Later, so-called "pick-up" method was reported[12,41], where exfoliated crystals with clean surfaces are mechanically assembled without exposing the interfaces to other wet-chemicals. Due the strength of the technique to form pristine interfaces, it has become a standard method to fabricate various artificial assemblies. Then, to increase the speed of controlled assembly, automated stacking was developed[42]. However, the assembly yield is still limited in this approach by low production of the assembly units. Based on CVD-grown semiconducting TMDs, the development has been made with extended scalability by combining the crystal growth and clean assembly techniques, listed above[2,36,43-45]. On the other hands, for graphene and hBN, the key two-dimensional materials, have been grown by CVD processes on large-scale[14,15,46-49], the atomically clean assembly has not been developed before our report.

### XRD from twisted graphite

XRD measurements compared the interfaces of two types of twisted graphite, one prepared using our PCA method, and one prepared using conventional wet-transferred assembly. Both samples with $\theta_i = 20°$, $N_L = 10$, showed clear diffraction peaks, which was narrower and more intense in the sample that had been fabricated by the PCA method than in the one prepared using wet-transferred assembly. The averaged $d_z$ were deduced from the peak positions, as 3.42 Å for the PCA sample, that is consistent with a theoretical value for pristine interfaces[50], and 3.50 Å for the wet-transferred sample, roughly 2 % higher than $d_z$ for the pristine interfaces. The thickness of coherently reflecting lattices was estimated using the Scherrer equation, $K \cdot \lambda / (\beta_{2theta} \cdot \cos(theta))$, where $K$ is a shape factor of 0.89[51], and $\beta_{2theta}$ is full width at half maximum of the $2theta$ peak. The coherent thicknesses were estimated as 36.7 Å and 27.8 Å, respectively for the two samples. Significantly, the estimated value from the PCA sample is close to the total thickness of the film; this result further confirms the consecutive formation of pristine interfaces to yield effective constructive interferences.

The $d_z$ were further measured using XRD for CTG with different $\theta_i$ and $N_L > 6$ to detect sufficient reflecting signals. Twisted interfaces resulted in quasi-periodic Moiré supercells, within which the atomic stacking configurations locally vary. In particular, so-called AA stacked regions with the overlapped two hexagonal lattices are mainly responsible for the increase in average $d_z$, with high local $d_z$. At small $\theta_i < \sim 3°$, spontaneous strong structural relaxation can maximize the area of energetically stable AB regions with Bernal-configuration, thereby minimizing the area of AA stacked regions[5]. As $\theta_i$ increases from 0°, AB regions with atomic registry are transformed to AA regions by the applied shear stress, thereby increasing the average $d_z$ until the transformation completes near $\theta_i = 3.3°$[5]. After the full transformation, the films become much flattened with weak relaxations. Considering the local variations of $d_z$, the theoretical values for averaged $d_z$ were identified (Figure 2f); the calculation assumes a single interface, so the actual value for $d_z$ in multilayers could be higher than the one for a single interface, depending on how the interfaces of small corrugations are overlapped with lateral translations, especially at small $\theta_i$. For the sake of simplicity, we did not consider this effect here.

**ARPES data from tri-layer CTG and ATG**

The hybridized electronic states that are produced by interlayer interactions were compared between tri-layer CTG and ATG by ARPES measurements. We defined the hybridization points (HPs) in momentum space, where the hybridized bands lead to the biggest mini-gap. Band structures were first obtained along the line between two Dirac points of the interacting layers (fig. S15A, green line). The band structures were taken along the line cut, while moving it toward the Γ point to search for the HP. The band structure (fig. S15B) of CTG ($N_L = 3$, $\theta_i = 20°$) was obtained across the HP (fig. S15A, red dot). The formation of a mini-gap is clearly indicated in the intensity profile as a function of energy (fig. S15D) at the HP along the dotted line in fig. S15B. The size of the band gap between the two local maxima of the intensity is ~200 meV. Hybridization proceeds by quasi-periodic Moiré potentials at the twisted interfaces, to yield Moiré Brillouin zones (mBZs) at the momentum space by continuum models (fig. S15A, gray hexagons). At the similar position within the mBZ, additional HPs (fig. S15A, blue dot) can be formed by adiabatic umklapp scattering in the Moiré potentials[52]. The size of mini-gap at the additional HP was measured to be the same as the one at the primary HP (fig. S15C, D). For ATG, we measured the mini-gap at the replica HP (fig. S15E, F) to observe the hybridized parabolic bands clearly without the diagonally piercing linear band that was observed at the primary HP. The mini-gap from the intensity profile (fig. S15G) at the HP was measured as 300 meV, which is significantly higher than the one from CTG; this increase suggests that hybridization can happen across multiple interfaces to form new electronic states.

## Circular dichroism from multilayer CTG

Structurally-programmable band structures with effective interlayer interactions can host a range of physical properties that have unique band topology and symmetry[4,6,25–29]. One such phenomenon is its chiro-optical property, such as circular dichroism, which is absent in natural graphite. Interactions between two Dirac cones, twisted in momentum space, result in chiral hybrid states, which demonstrate electromagnetic coupling. Optical excitations of the states (fig. S17A, dashed lines) produce in-plane magnetic dipole moments, which are responsible for circular dichroism[53]. An increase in $N_L$ combined with appropriate $\theta_i$ could further strengthen this phenomenon and even result in new collective electronic states from multiple interfaces[28,29]. However, these electronic structures and properties are largely unexplored owing to the unavailability of a material-design tool to reliably fabricate pristine interfaces.

To demonstrate the programming of intrinsic properties in the multilayer films with large $N_L$, optical ellipticity $\Psi$ (or circular dichroism) spectra were recorded for CTG with a constant $\theta_i = 20°$ while increasing $N_L$, up to 6 (fig. S18A). $\Psi$ represents the difference in absorption $I_L$ of left-handed circularly polarized light and absorption $I_R$ of right-handed circularly polarized light, with respect to the receiver as $\Psi = (I_L - I_R)/[2(I_L + I_R)]$[53]. The multilayer-sample spectrum ($\theta_i \approx 0°$) did not show noticeable peaks (fig. S18A, gray spectrum), the CTG-sample spectra showed two dominant peaks, one at 3.28 and one at 4.54 eV. The energy levels are consistent with previously-reported interlayer optical-excitation energies for $\theta_i = 20°$, which are associated with the *A* and *B* transitions[54] (fig. S17A). Importantly, the spectral weights of the peaks increased with increasing $N_L$ (fig. S18B).

Electrical and optical properties related to the hybrid states are very sensitive to the interface quality because weak van der Waals interactions are effectively suppressed by contaminant layers that are even one atom thick. Therefore, the results indicate the formation of pristine interfaces and demonstrate the reliable structural programming of physical properties in multilayer van der Waals structures.

**Tunneling model for vertical transports across Gr/hBN/Au junctions**

The formula for tunneling current density with an applied bias $V$ at the Au electrode with respect to graphene[21,55] is

$$J(V) \propto \int \{f(E-eV) - f(E)\} DoS_{Au}(E-eV) DoS_{Gr}(E) D(E,V) dE, \quad \text{(Eq. 1)}$$

where $f(E)$ is the Fermi-Dirac distribution function, $DoS_{Au}(E)$ is the electronic density of states for Au, and $e$ is the elementary charge, $DoS_{Gr}(E)$ is the electronic density of states for graphene, and $D(E, V)$ is the tunneling probability. Here, we assume that the in-plane momentum is not conserved. Considering the Fermi functions, the relevant energy range for the integral is confined to $\mu - eV < E < \mu$ for the case of $eV > 0$, where $\mu$ is chemical potential of graphene. The $DoS_{Au}(E-eV)$ can be set as a constant, and $DoS_{Gr}(E)$ is described as $2|E|/(\pi\hbar^2 v_F^2)$, when the energy level of Dirac point is defined as 0, where $\hbar$ is the reduced Planck constant, and $v_F$ is Fermi velocity. Then Eq. 1 can be expressed as

$$J(V) \propto A \int_{\mu(V)-eV}^{\mu(V)} |E| D(E,V) dE, \quad \text{(Eq. 2)}$$

where the constant $A$ is described as $8m^*e/(\pi\hbar^4 v_F^2)$, where $m^*$ is the electron effective mass in Au. The two $V$-dependent terms of $\mu(V)$ and $D(E, V)$ can be described as follows.

First, $\mu(V)$ of graphene with limited $DoS$ changes as a function of $V$ by the capacitive coupling with Au through hBN tunneling barriers. Therefore,

$$\mu(V) = \pm \hbar \nu_F \sqrt{\pi \frac{\varepsilon_{hBN}}{t_{hBN}} \frac{|V - V_D|}{e}}$$ (+ for $V > V_D$, − for $V < V_D$), where $\varepsilon_{hBN}$ and $t_{hBN}$ are respectively

the dielectric constant and the thickness of hBN, and $V_D$ is the applied bias to align the Fermi level of graphene to the Dirac point.

Second, $D(E, V)$ is a function of both $E$ and $V$. Here, the hBN tunneling barriers is treated as an isotropic material with a standard theory for tunneling toward the out-of-plane direction; this assumption has been confirmed to be valid[21]. We conclude that holes are the dominant carriers for the tunneling, as discussed in the next paragraph. Then,

$$D(E, V) \propto exp\left( -\frac{2t_{hBN}}{\hbar} \sqrt{2m_{hBN}^* \left( \phi_{B0} + E + \frac{eV + \Delta W + \mu(V)}{2} \right)} \right)$$, where $m_{hBN}^*$ is the effective hole

mass in hBN toward the tunneling direction, $\Delta W$ is work function difference between Au and graphene, and $\phi_{B0}$ is the tunneling barrier height from the Dirac point, to which Fermi level is aligned. Here, $\phi_{B0} + (eV + \Delta W + \mu(V)) / 2$ indicates the average barrier height $\phi_B$.

For the simulation of d$I$/d$V$-$V$ characteristic of the Figure 3d in the main manuscript, we numerically calculated Eq. 2 with $\varepsilon_{hBN} = 2.17$[56], $V_D$, measured from the dip in the d$I$/d$V$-$V$ curve in positive bias voltage (Figure 3d), and $\phi_{B0}$ deduced from $t_{hBN}$-dependent $R_0A$ (see following discussion). From the simulated $J$-$V$ (fig. S13A), we deduced the d$I$/d$V$ curve. The resulted d$I$/d$V$ curve

was smoothed, considering any inhomogeneity across the junction area. An energy threshold of

±63 meV at $V = 0$ was introduced; it was defined by the phonon energy to mediate the tunneling[23],

which is responsible for the dip of d$I$/d$V$ near $V = 0$. The simulated result agreed well with the

experimental data, and successfully reproduced the observation, where two local minima of d$I$/d$V$

appear, one at $V = V_A$ and one at $V = V_D$. In particular, our theoretical modeling suggests that holes are majority carriers for tunneling. When we assumed that the electron is the majority carrier, the simulated $J$-$V$ curve was highly asymmetric with a much lower $J$ at $V < 0$ than $J$ at $V > 0$, because the strongly negative $V$ lowers the $\mu(V)$ of graphene from the edge of conduction band of hBN,

and thereby decreases $D(E, V)$ by the increased barrier height for electron. In contrast, $D(E, V)$

increased with increase in the change of $\mu(V)$ for hole transport. Our experimental data (fig. S13B)

are consistent with the latter case because of the higher $J$ in the negative $V$ region than in the positive $V$ region.

**Estimation of tunneling barrier height**

We estimated $\phi_{B0}$ from $t_{hBN}$-dependent $R_0A$ (Figure 3b, inset in the main manuscript). From Eq. 1, $R_0A$[21,55] is derived as

$$R_0A \propto 1 / \left( \hbar\sqrt{\phi_B} / (2t_{hBN}\sqrt{2m_{hBN}^*}) + |\mu(0)| \right) e^{\frac{2t_{hBN}}{\hbar}\sqrt{2m_{hBN}^*\phi_B}}, \qquad \text{(Eq. 3)}$$

where $\mu(0)$ is the initial chemical potential of graphene at the zero-bias, which was estimated as $-0.4$ eV for Figure 3d with the measured $V_D$, and $m_{hBN}^* = 0.5\,m_0$, where $m_0$ is the mass of a free electron. When $t_{hBN}$ varies, the $R_0A$ is mostly affected by the exponential component in Eq. 3, and therefore shows an exponential increase that depends on $\phi_B$. The fitting (Figure 3b inset, dotted line) by taking the $\phi_{B0}$ within the $\phi_B$ as a free parameter, results in a value of 3.0 eV for hole transport; this value is similar to the difference between the Fermi level and valance band maximum of hBN on graphite, measured by ARPES[22].

monolayer MoS$_2$ and WS$_2$ by photoluminescence excitation spectroscopy. *Nano Lett.* **2015**, *15*, 2992–2997.

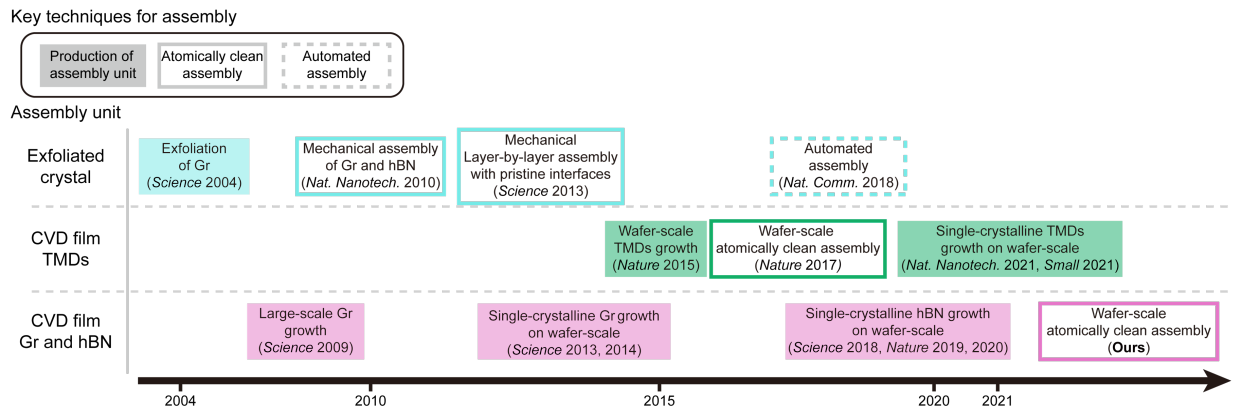

Fig. S1. Summary of 2D assembly developments

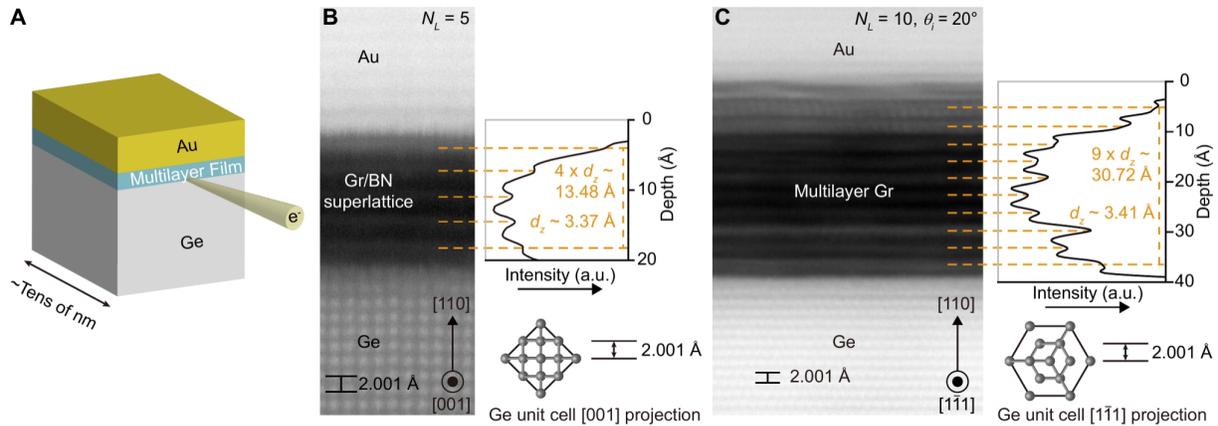

**Fig. S2**. **Cross-sectional STEM images of multilayer films.** (**A**) Schematic for cross-sectional STEM. (**B, C**) Annular dark-field STEM image (left) and the vertical intensity profile (right) for graphene/hBN superlattice film ($N_L$ = 5, B) and CTG ($N_L$ = 10, $\theta_i$ = 20°, C) on Ge(110) substrates with Au capping layers. Taking the inter-atomic distance along the [110] direction in Ge as a reference, the interlayer distances between graphene and hBN layers, and graphene layers are estimated as 3.37 Å and 3.41 Å, respectively.

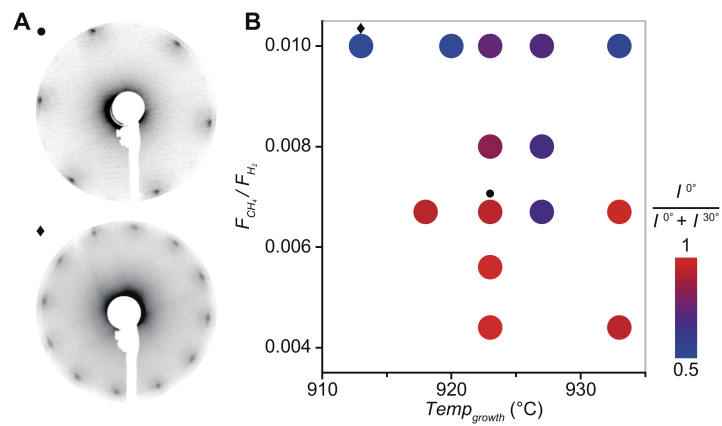

**Fig. S3. Growth condition–dependent, in-plane crystalline orientations of graphene.** (**A**) Representative LEED patterns of graphene films grown under two different growth conditions, which are indicated by symbols in (B). The top image (circle) shows a single hexagonal pattern, but the bottom image (diamond) shows two dominant overlaying hexagonal patterns, one rotated by 30° with respect to the other. (**B**) Color-coded map of intensity ratio, $I^{0°}/(I^{0°}+I^{30°})$, obtained under different growth conditions, where $I^{0°}$ and $I^{30°}$ are the integrated intensities for the major and minor diffraction spots, which are rotated by 30°. The x-axis is growth temperature, and the y-axis is the ratio of the flow rates of $CH_4$ and $H_2$ ($F_{CH_4}/F_{H_2}$). Two crystalline orientations are preferred, and low $F_{CH_4}/F_{H_2}$ is generally required to obtain a single-crystal orientation, as previously reported[14].

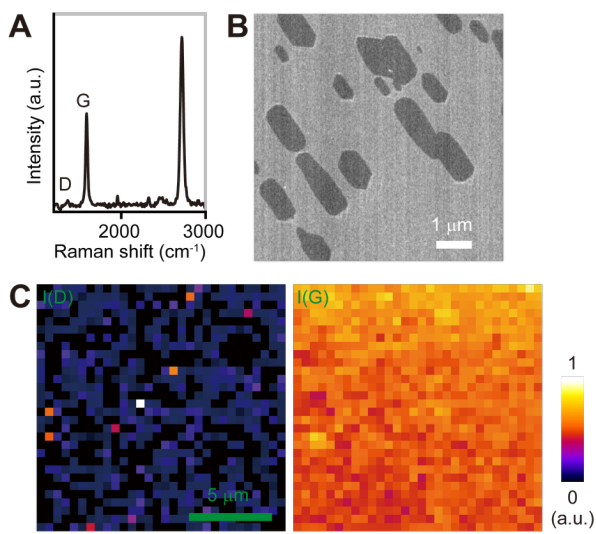

**Fig. S4. Quality of as-grown graphene films.** (**A**) Raman spectra of single layer graphene transferred on SiO₂/Si substrates. It shows no significant defect-associated D peak near 1,350 cm⁻¹. (**B**) Scanning electron microscope image of partially grown graphene on Ge substrate. The averaged sizes of individual domains are 1~3 μm wide. (**C**) Raman intensity maps for D and G peaks across an area, where multiple domains merged. The negligible D peak signals confirm that multiple domains merge into high-quality films.

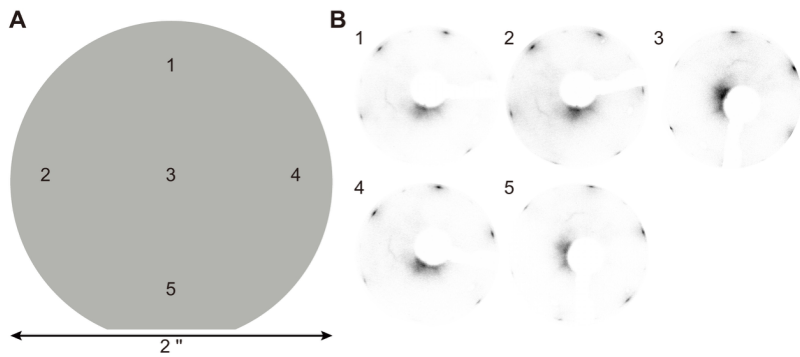

**Fig. S5. In-plane crystalline orientation of graphene on Ge(110). (A)** Schematic of a 2" Ge(110) wafer. **(B)** LEED patterns of graphene at five representative areas across the 2" growth zone. LEED patterns indicate that the graphene has a uniform crystallographic orientation.

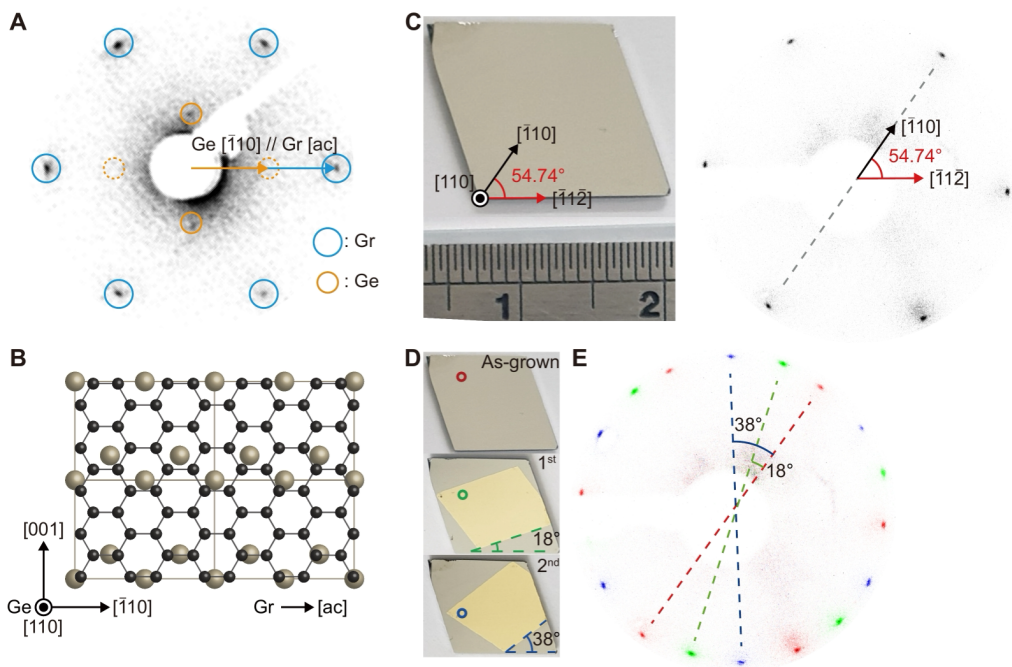

**Fig. S6. Twist angle-controlled stacking process.** (**A**) LEED pattern of unidirectionally-aligned graphene on Ge(110). Spots encircled in blue and orange are graphene and Ge diffraction signals, respectively. The armchair direction of graphene is parallel to the Ge[$\bar{1}$10] orientation. (**B**) Schematic of the most energetically-favorable atomic configuration of epitaxially-grown graphene on Ge(110)[57]. (**C**) (left) Photograph of as-grown graphene on Ge(110) substrate, and (right) corresponding LEED pattern. The Ge[$\bar{1}$10] direction can be identified by reference to the straight edge of the Ge[$\bar{1}$1$\bar{2}$] crystalline orientation. (**D**) Photographs after consecutive transfers of graphene onto as-grown substrate. Top: as-grown sample, middle: after 1$^{st}$ transfer of graphene, bottom: after 2$^{nd}$ transfer of graphene. Here, each graphene layer was intentionally stacked with a supporting Au superlayer to increase visibility of the edge. (**E**) Combined LEED patterns with false colors. Each colored pattern is separately measured corresponding to the areas encircled in (D). The twist angles between the patterns match the relative orientations of the Ge[$\bar{1}$1$\bar{2}$] edges; this result confirms the success of edge-based twist-angle control.

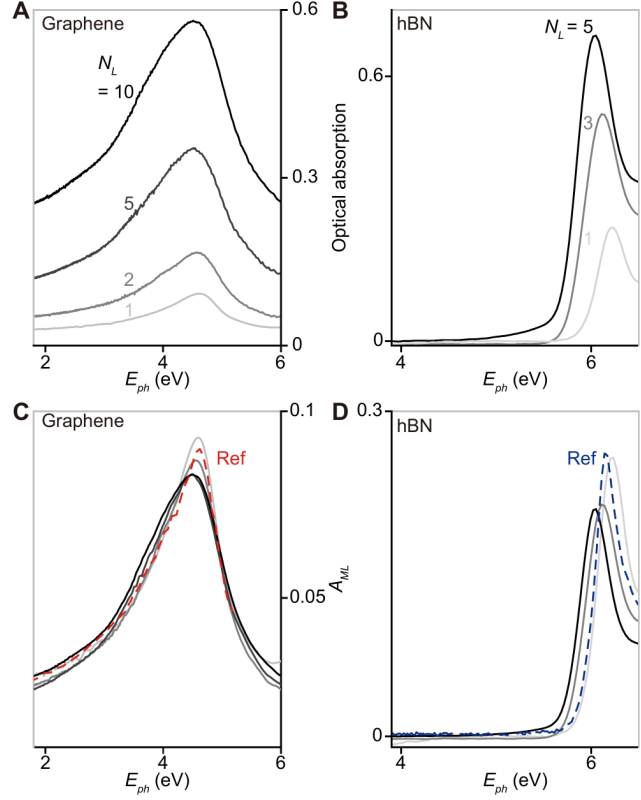

**Fig. S7. Near-unity stacking yield for graphene and hBN.** (**A**, **B**) Optical absorption spectra for multilayer graphene (A) and hBN (B) with different $N_L$. (**C**, **D**) Optical absorption for ML, $A_{ML}$, deduced from the data for multilayer graphene (C) and hBN (D) films with different $N_L$. Deduced $A_{ML}$ spectra are consistent with reference data[58] (dotted line), and similar for different $N_L$; this result confirms the near-unity stacking yield.

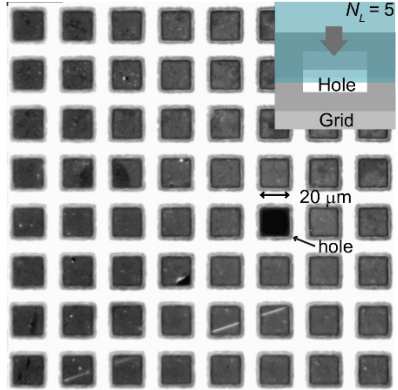

**Fig. S8. Mechanical robustness of multilayer graphene membrane.** Optical image of multilayer graphene ($N_L$ = 5) suspended on a holey grid (each hole is 20 μm × 20 μm). Arrow: an area without a film as a reference. The data illustrates the mechanical strength of the ultra-thin film.

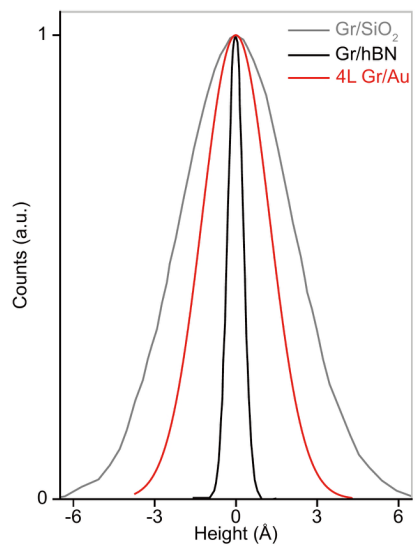

**Fig. S9. Surface flatness of multilayer graphene.** Graph of the height distribution for the stacked graphene ($N_L$ = 4, red line), graphene on hBN[59] (black line), and graphene on $SiO_2$[59] (gray line) in 100 nm × 100 nm area. Stacked graphene shows a narrower height distribution than graphene on $SiO_2$.

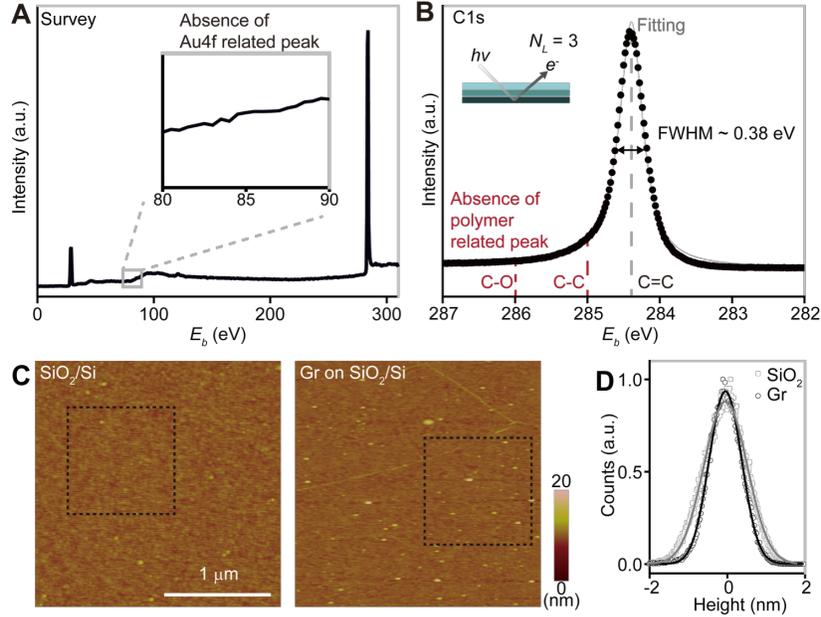

**Fig. S10. Surface characterization on transferred graphene.** (**A**) XPS spectra on transferred multilayer graphene ($N_L = 3$) after etching the Au supporting layer. (**B**) C1s peak, fitted with the known asymmetric peak for carbons with a graphitic $sp^2$ bond. Other peaks, associated with C-O and C-C bonds are absent. Inset: schematic for XPS measurement on a multilayer graphene film. The spectrum shows a sharp peak with smaller FWHM of 0.38 eV than as-grown graphene on SiC substrates (FWHM = 0.68 eV). (**C**) AFM height images on a SiO₂/Si substrate before (left) and after graphene transfer (right). (**D**) Histograms of heights, taken from the dotted regions of each sample in (C), showing similar distributions

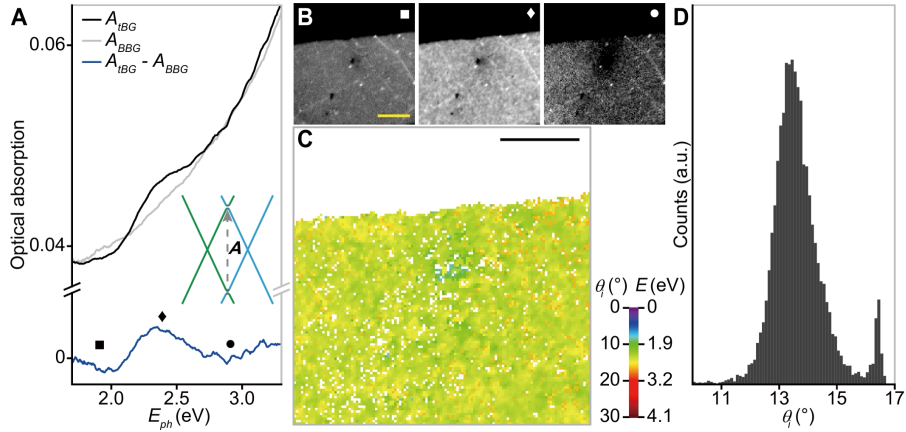

**Fig. S11. Interlayer optical absorption of twisted bilayer graphene film**. (**A**) Optical absorption spectra for twisted $A_{tBG}$ (black line) and Bernal-stacked bilayer graphene $A_{BBG}$ (gray), measured over a 50 μm × 75 μm area. $\theta_i = 13.5°$ for the twisted sample. The difference $A_{tBG} - A_{BBG}$ (blue) shows an additional absorption peak near 2.38 eV, which corresponds to the interlayer optical transition **A** between van Hove singularities in twisted bilayer graphene with $\theta_i = 13.5°$ for pristine interfaces (inset). (**B**) Optical images of $A_{tBG} - A_{BBG}$ at different photon energies, indicated by symbols in a (from left to right, 1.9, 2.38 (±0.1), and 2.9 eV). Absorption increases near the interlayer transition energy with spatial uniformity. (**C**) Map of $\theta_i$, estimated from the $\theta_i$ dependent absorption peak energies for $A_{tBG} - A_{BBG}$[39], which shows uniform $\theta_i$. (**D**) Histogram of $\theta_i$ extracted from 50 μm × 75 μm area in (C). All scale bars: 20 μm.

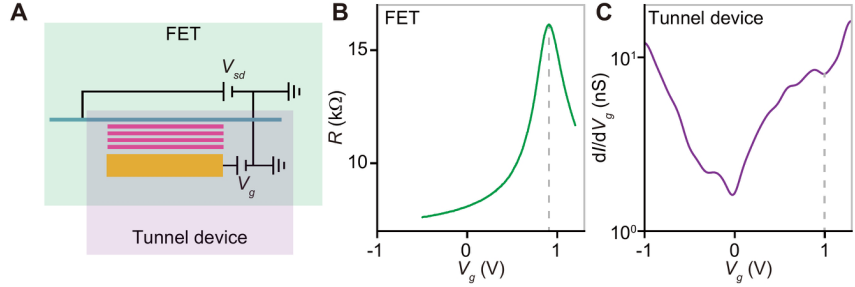

**Fig. S12. Gr/hBN/Au junction devices.** (**A**) Schematic of measurement geometry for vertical transport (purple box, tunnel device geometry) and lateral transport (green box, field-effect transistor geometry). (**B**) In-plane resistance of the graphene electrode as a function of the gate voltage $V_g$ applied to the Au bottom electrode. Dashed line: $V_g$ to align the Fermi level of graphene to the Dirac point. Here, $V_{sd}$ was set to 0.1 V. (**C**) d$I$/d$V_g$ curve measured across the Gr/hBN/Au junctions. It shows a local minimum at the Dirac point voltage in (B).

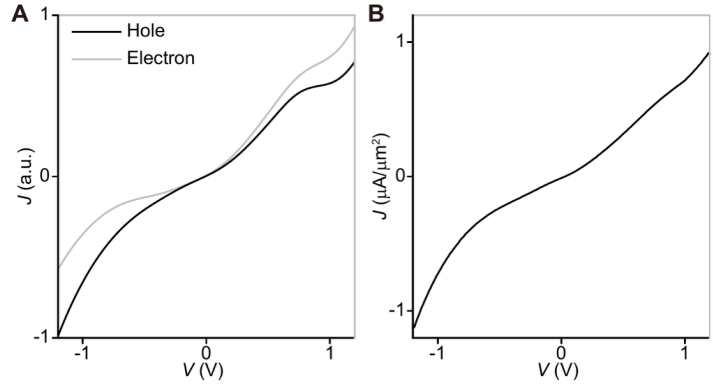

**Fig. S13. *J-V* characteristics across Gr/hBN/Au junctions.** (**A**) Simulated curve with a tunnelling model, assuming hole (black line) and electron (gray line) transports. (**B**) Experimentally-measured *J-V* curve for tunnel device in Fig. 3b ($N_L = 3$).

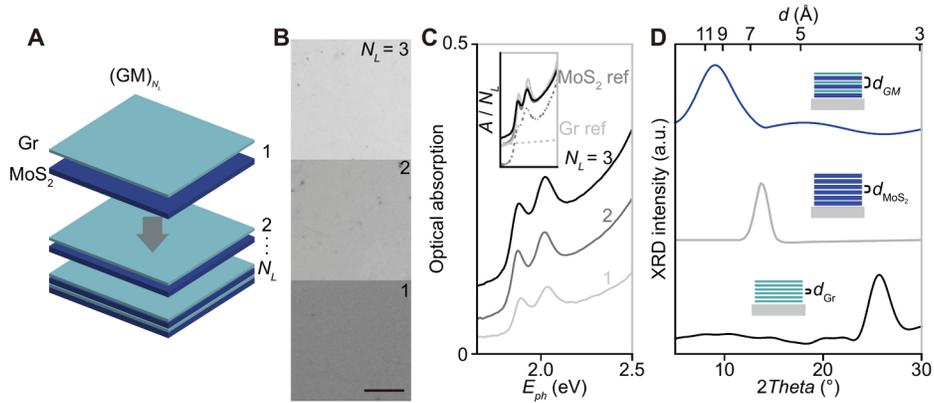

**Fig. S14. Fabrication of graphene/MoS₂ superlattices.** (**A**) Schematic of a graphene/MoS₂ superlattice prepared by stacking graphene/MoS₂ ML units $N_L$ times. The graphene and MoS₂ layers were mechanically assembled. (**B**) Optical reflectance image of superlattice films with $N_L$ = 1, 2, and 3, respectively, on fused silica substrates. Scale bar: 100 μm. (**C**) Optical spectra of the superlattices with $N_L$ = 1, 2, and 3. Inset: normalized spectra with reference data for graphene[58] and MoS₂ MLs[60] (dashed lines). (**D**) Top to bottom: XRD patterns for the graphene/MoS₂ superlattice, multilayer MoS₂, and multilayer graphene, respectively. $N_L$ for the films was determined, to achieve a total thickness of ~3 nm ($N_L$ = 3, 4, and 10 for the superlattice, MoS₂, and graphene films, respectively).

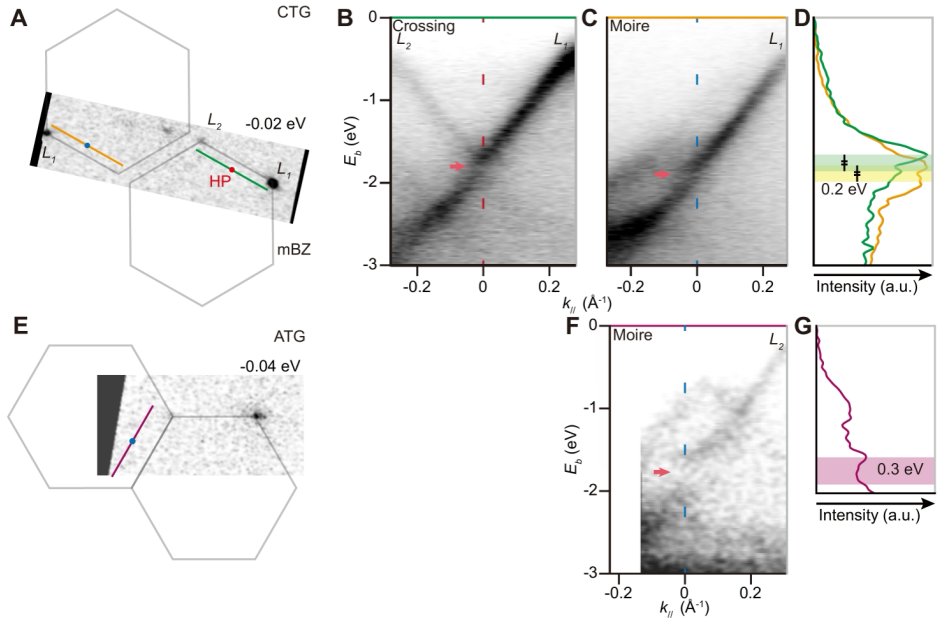

**Fig. S15. ARPES data from CTG and ATG.** (**A**) Photoemission contour in CTG ($N_L = 3$, $\theta_i = 20°$) at electron binding energy $E_b = -0.02$ eV. Hexagons: mBZ by Moiré potentials at the twisted interface between $L_1$ and $L_2$ layers. Red dot: primary HP, where two bands from $L_1$ and $L_2$ layers intersect, blue dot: replica of HP by periodic Moiré potentials. (**B**, **C**) Band structures taken along the green line (B) in (A), across the HP, and the orange line (C) in (A), across the replica of HP. (**D**) Photoemission intensity profiles along the dashed lines in (B) (green) and (C) (orange). Intensity dips (highlighted in green and yellow) indicate mini-gaps of ~0.2 eV. (**E**) Photoemission contour in ATG ($N_L = 3$, $\theta_i = \pm20°$) at $E_b = -0.04$ eV. Hexagons: extended mBZ, together with the replica of HP between $L_1$/$L_3$ and $L_2$ bands. (**F**) Band structure taken along the purple line in (E). (**G**) Photoemission intensity profile along dashed lines in (F), presenting a mini-gap of ~0.3 eV.

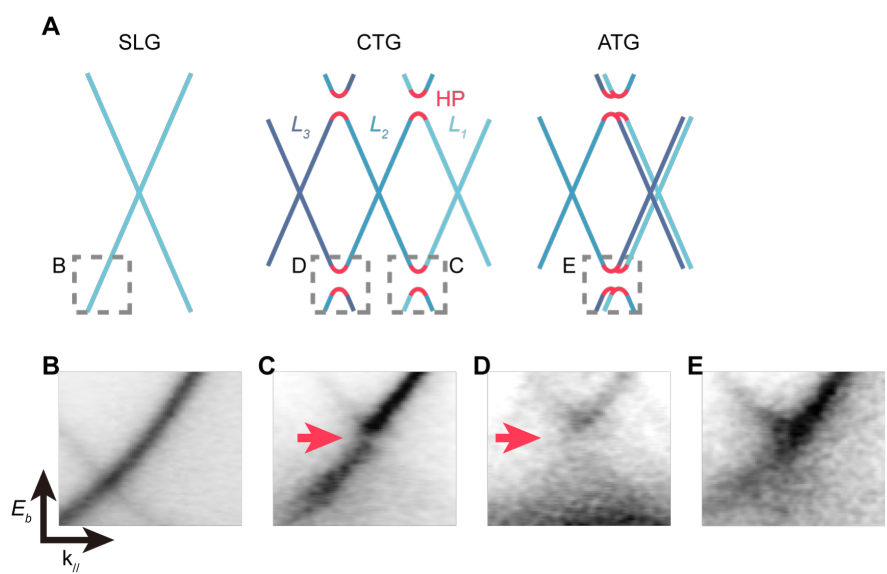

**Fig. S16. Mini-gaps at HPs** (**A**) Band structure for SLG, CTG and ATG (from left to right) (**B-E**) ARPES data for the dotted regions in (A). The arrows indicate the mini-gaps at the HPs.

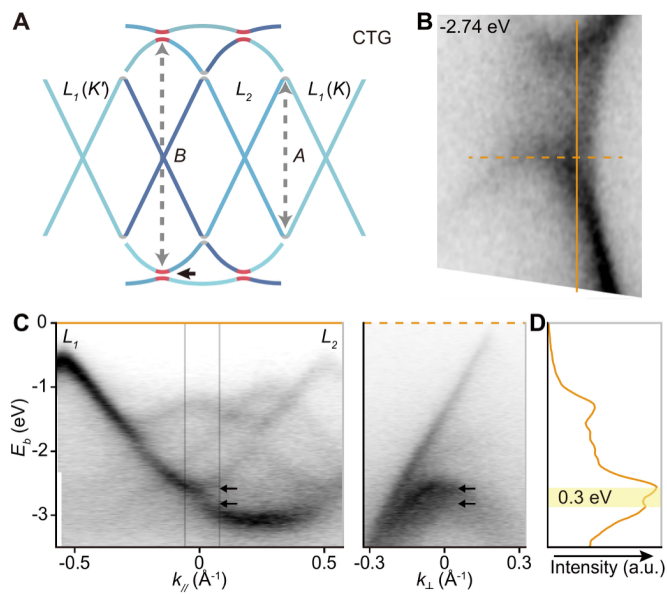

**Fig. S17. Hybrid electronic states associated with interlayer interactions in multilayer graphene.** (**A**) Schematic of band structure for CTG, color-coded bands represent different layers. Dirac cones from the nearest layers form hybridized states (red) with mini-gaps, associated with interlayer optical transition **_B_**. (**B**) Photoemission contour map at $E_b = -2.74$ eV near the hybridized states in (A). (**C**) Band structures taken along the orange lines in (B), across two Dirac cones from $L_1$ and $L_2$ layers. (**D**) Cumulative photoemission intensity spectrum within the gray box region in (C).

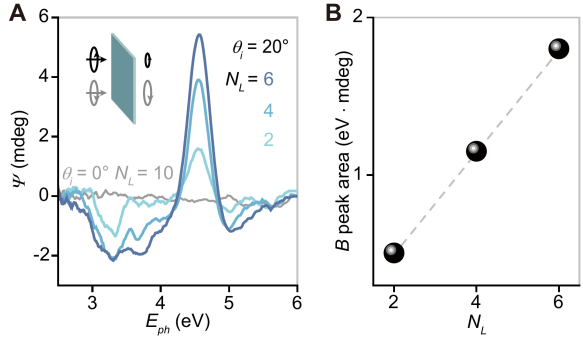

Fig. S18. Chiro-optical properties in CTG. (A) Optical ellipticity $\Psi$ spectra of multilayer films with $N_L = 10$, $\theta_t \approx 0°$ (gray) and $N_L = 2\sim6$, $\theta_t = 20°$ (cyan). (B) Integrated areas of $\Psi$ peaks associated with the transition $B$ in Fig. S17A, as a function of $N_L$ for samples with $\theta_t = 20°$.